\documentstyle[aps,prl,multicol,floats,epsf]{revtex}

\begin{document}

\draft

\wideabs{

\title{Unusual magnetic susceptibility anisotropy in untwinned
La$_{2-x}$Sr$_x$CuO$_4$ single crystals in the lightly-doped
region}

\author{A. N. Lavrov, Yoichi Ando, Seiki Komiya, and I. Tsukada}

\address{Central Research Institute of Electric Power Industry,
2-11-1 Iwato-kita, Komae, Tokyo 201-8511, Japan}

\date{\today}
\maketitle

\begin{abstract}

We present a study of the magnetic susceptibility $\chi$ in carefully
detwinned La$_{2-x}$Sr$_{x}$CuO$_4$ single crystals in the lightly-doped
region ($x=0-0.03$), which demonstrates a remarkable {\it in-plane}
anisotropy of the spin system. This anisotropy, $\chi_a/\chi_b$, is
found to persist after the long-range antiferromagnetic (AF) order is
destroyed by hole doping, suggesting that doped holes break the AF order
into domains in which the spin alignment is kept essentially intact. It
turns out that the freezing of the spins taking place at low
temperatures is also notably anisotropic, implying that the
``spin-glass" feature is governed by the domain structure as well.

\end{abstract}

}

It is generally believed that the interplay between doped holes and
antiferromagnetic (AF) correlations governs the occurrence of high-$T_c$
superconductivity in cuprates. To clarify the mechanism of
superconductivity, it is therefore indispensable to thoroughly
understand the environment for the hole motion, and thus the magnetic
structure of CuO$_2$ planes and its evolution with doping are of
particular importance.

Parent insulating cuprates at high temperatures have been considered to
be relatively simple: they are essentially two-dimensional (2D)
Heisenberg antiferromagnets \cite{review,Keimer} and a weak interplane
interaction gives rise to the three-dimensional (3D) long-range N\'{e}el
order \cite{review,Keimer}. However, the doped holes significantly
complicate the picture by introducing frustrations in the spin system;
the origin of the frustration has been discussed in the literature
\cite{SG1,vortex,texture}, but it is not well understood yet. Existing
theories for the lightly-doped region \cite{SG1,vortex,texture}
conjecture that the doped holes are distributed homogeneously and cause
spin distortions that spread more than several unit cells in the CuO$_2$
planes; this picture results in a fluctuating short-range AF order at
high temperatures (where the holes move) \cite{vortex} and some kind of
spin-glass (SG) state at low temperatures (where the holes localize)
\cite{SG1,texture}. The spin-glass features observed in twinned
La$_{2-x}$Sr$_{x}$CuO$_4$ (LSCO) crystals \cite{SG2,SG3} have been
discussed to support this picture. A salient point of this homogeneous
picture is that once doped holes destroy the long-range AF order, the
spin system should become essentially {\it isotropic} in the CuO$_2$
planes. However, recent neutron diffraction studies of LSCO
\cite{1Da,1Db} have found a one-dimensional spin modulation throughout
the SG doping range \cite{1Da,1Db}, which casts serious doubt on the
isotropic scenario. The observed one-dimensional modulation naturally
implies that the hole doping forces the long-range AF order to cross
over into a unidirectional spin-density wave (SDW), presumably
associated with charge ordering \cite{stripe,ourstr}, and the system can
no longer be homogeneous.

In order to clarify the true picture for the evolution of the magnetic
order upon hole doping, the magnetic susceptibility $\chi$ in
lightly-doped LSCO is useful. The orthorhombic structure in LSCO gives
rise to the Dzyaloshinskii-Moriya (DM) interaction, and the spins are
weakly canted from the direction of the staggered magnetization because
of the DM interaction \cite{MR_WF,peak,Tsukada}. This spin canting
causes weak ferromagnetic moments to be associated with the AF
correlations in LSCO, and one can use these moments as a probe to trace
the evolution of the magnetic order \cite{peak}.

In this Letter, we present a detailed study of the static magnetic
susceptibility in detwinned La$_{2-x}$Sr$_{x}$CuO$_4$ ($x=0-0.03$)
single crystals. In all the samples, the susceptibility is found to
exhibit unusual anisotropies, particularly within the CuO$_2$ planes, in
the ``paramagnetic" state, which is not expected for a 2D Heisenberg
antiferromagnet. The data show that the spin alignment remains
anisotropic even in the absence of long-range AF order, which indicates
that doped holes break the AF order into domains or replace it with a
spin-density wave rather than introduce mesoscopically-homogeneous spin
distortions. Correspondingly, the low-temperature spin freezing in our
untwinned crystals is found to be anisotropic (with anisotropic Curie
constant and anisotropic SG temperature $T_{SG}$), which calls for a
fundamental revision of the standard picture for the spin-frozen state
in cuprates.

\begin{figure}[!ht]
\leftskip12pt
\vspace{8pt}
\epsfxsize=0.79\columnwidth
\centerline{\epsffile{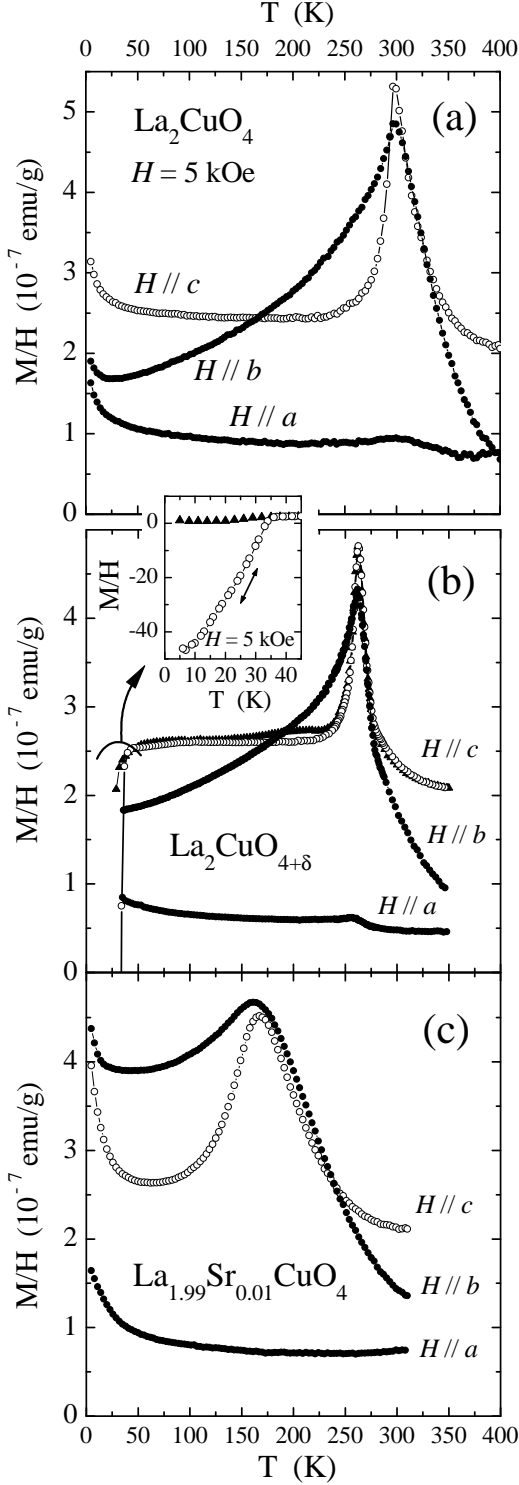}}
\vspace{10pt}
\caption{Magnetic susceptibility of untwinned single crystals of (a)
La$_{2}$CuO$_4$, (b) La$_{2}$CuO$_{4+\delta}$, and (c)
La$_{1.99}$Sr$_{0.01}$CuO$_4$, measured in $H$=5 kOe applied along one
of the crystal axes. In panel (b), different symbols for $H\parallel c$
correspond to data taken on heating after quick (solid triangles) and
slow (open circles) cooling from 300 to 5 K; Inset illustrates the
impact of cooling conditions on the occurrence of superconductivity.}
\label{fig1}
\vspace{-22pt}
\end{figure}

High-quality La$_{2-x}$Sr$_x$CuO$_4$ ($x$ = 0, 0.01, 0.02, and 0.03)
single crystals are grown by the traveling-solvent floating-zone (TSFZ)
technique. Repeated X-ray analyses are employed to prepare
parallelepiped samples (typical weight is $\sim$40 mg), where all the
faces are adjusted to the orthorhombic crystal planes with an accuracy
of 1$^{\circ}$. In order to remove excess oxygen, the samples are
carefully annealed at 350 -- 680$^{\circ}$C in a flow of pure helium.
For the compositions in this study, LSCO undergoes a transition from
high-temperature tetragonal to low-temperature orthorhombic phase at 200
-- 250$^{\circ}$C \cite{Keimer}; this transformation usually results in
heavy twinning that makes the $a$ and $b$ orthorhombic axes to be
macroscopically indistinguishable (we use the notation where the $c$
axis is perpendicular to CuO$_2$ planes). To obtain untwinned crystals,
we slowly cool the samples under a uniaxial pressure of 15 -- 30 MPa
from 270$^{\circ}$C. According to X-ray measurements, the resulting
fractions of misoriented domains (which characterize the quality of
detwinning) are $\leq 1\%$ in the $x=0.01$ sample, $\sim$2.5\% in the
$x=0.03$ sample, and $5-10\%$ in the rest, depending presumably on the
crystal defect structure. It is found that the magnetic susceptibility
is also an excellent tool for checking the detwinning ratio; while the
X-ray probes only the surface, the in-plane susceptibility anisotropy
$\chi_a/\chi_b$ gives a good measure of the detwinning in the bulk, and
the fraction of misoriented domains in our samples estimated this way is
below 5\%. Magnetization measurements are carried out using a Quantum
Design SQUID magnetometer at fields from 0.2 to 5 kOe applied along one
of the orthorhombic crystal axes.

Figure 1(a) shows the anisotropic susceptibility of undoped
La$_{2}$CuO$_4$, where $\chi_b(T)$ and $\chi_c(T)$ demonstrate
pronounced peaks at the N\'{e}el temperature $T_N \approx 300$ K while
$\chi_a$ remains virtually unchanged through $T_N$ (a weak hump in
$\chi_a(T)$ is caused presumably by some admixture of $\chi_b$). The
$\chi_c$ data agree well with previous studies
\cite{MR_WF,peak,Tsukada}, where the peak in $\chi_c(T)$ has been
explained \cite{MR_WF,peak} to originate from the canted moments due to
the DM interaction. Note that the DM interaction confines the canted
moments in the $bc$ plane \cite{MR_WF,peak,Tsukada}. The contribution of
the canted moments to $\chi_c$ grows together with the AF correlation
length $\xi_{\rm AF}$, and it decreases abruptly when the canted moments
order antiferromagnetically along the $c$ direction below $T_N$. Note
that the data in Fig. 1(a) are consistent with the previous assertion
that the spin easy axis is $b$ and the canted moments are along $c$.

The behavior of $\chi_a$ and $\chi_b$ in Fig. 1(a), however, differs
significantly from what is expected for the suggested spin Hamiltonian
\cite{MR_WF,peak,Tsukada} which includes the isotropic Heisenberg term
and the DM term. The following unexpected features are noted: (i)
Although $\chi_b$ represents the transverse susceptibility of the canted
moments, it shows a marked decrease below $T_N$ instead of being
$T$-independent. (ii) $\chi_b(T)$ and $\chi_c(T)$ deviate from each
other above $\sim$330 K, indicating that $b$ and $c$ are nonequivalent
even in the ``paramagnetic" state. (iii) Upon transition from the 2D
Heisenberg to the 3D N\'{e}el state, the susceptibility should increase
(decrease) when it is measured transverse to (along) the spin easy axis
\cite{Manousakis}; therefore, a clear feature should be observed in
$\chi_a(T)$ at $T_N$. The absence of any notable anomaly in $\chi_a(T)$
across $T_N$ means that the spin directions are kept confined in the
$bc$ plane above $T_N$. The features (ii) and (iii) reveal that the
suppression of the long-range N\'{e}el order in undoped LSCO does not
make the spin system isotropic.

\begin{figure}[!t]
\vspace{1pt}
\leftskip15pt
\epsfxsize=0.74\columnwidth
\centerline{\epsffile{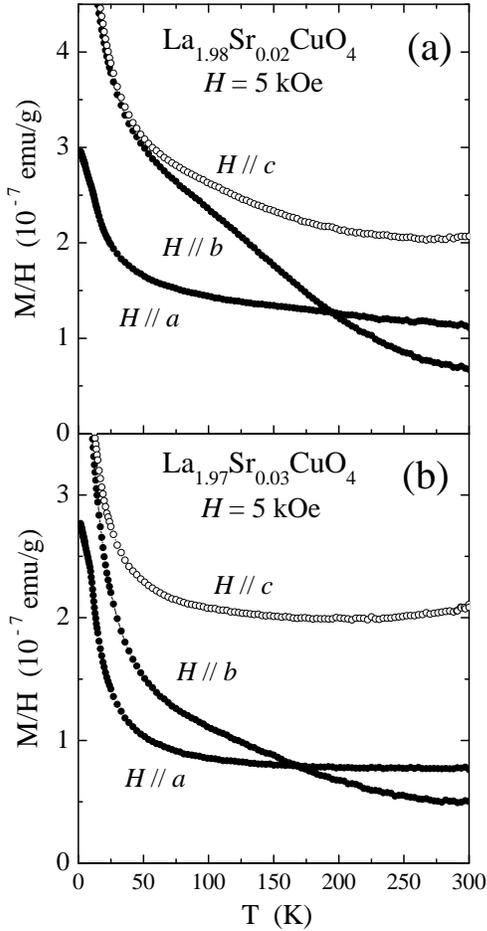}}
\vspace{6pt}
\caption{Magnetic susceptibility of La$_{2-x}$Sr$_{x}$CuO$_4$ with
(a) $x$=0.02 and (b) $x$=0.03.}
\vspace{-5pt}
\label{fig2}
\end{figure}

A common concern in studying lightly-doped LSCO is the oxygen
non-stoichiometry \cite{SG2,Tsukada}; to find out how it can affect the
susceptibility behavior, we have repeated measurements of the
La$_{2}$CuO$_4$ sample after adding some excess oxygen $\delta \leq
0.01$ [Fig. 1(b)]. The excess oxygen actually introduces new features
associated with the phase separation into macroscopic oxygen-rich and
oxygen-free regions \cite{separ}; when cooled slowly, the
La$_{2}$CuO$_{4+\delta}$ sample exhibits superconductivity with $T_c\sim
35$ K [inset of Fig. 1(b)], which comes from the oxygen-rich region. The
change of $\chi(T)$ around $T_N$ looks more complicated, and a decrease
in $T_N$ is accompanied by a sharpening of the peak in $\chi(T)$;
however, this sharpening is understood to be an artifact \cite{note} of
the phase separation taking place at $\sim$270 K. Hence, the features
that may be brought about by the oxygen non-stoichiometry can be
identified, and we can conclude that none of the susceptibility
peculiarities appearing in Fig. 1(a) is associated with excess oxygen.

Figure 1(c) shows the data for $x$=0.01, where only 1\% of hole doping
notably suppresses $T_N$ and broadens the corresponding peaks in
$\chi(T)$. Still, similarly to the undoped sample, $\chi_b(T)$ and
$\chi_c(T)$ deviate from each other at $\sim$50 K above $T_N$, and
$\chi_a(T)$ stays flat through $T_N$ (this sample is $\approx$100\%
detwinned). Apparently, there is no sign that an isotropic spin state is
recovered above $T_N$.

The unusual anisotropy is not confined in the AF region of doping, but
is found to persists to higher doping (Fig. 2); this is especially
surprising in view of the common idea that the CuO$_2$ planes become
nearly isotropic once the hole doping destroys the long-range AF order
\cite{review,Keimer,SG1,vortex,texture,SG2}. In both the $x$=0.02 and
0.03 samples, $\chi_b(T)$ and $\chi_c(T)$ behave essentially the same
way as in the antiferromagnetic samples ($x$=0 and 0.01) as if $T_N$=0,
which strongly suggest that the $T$-dependence of $\chi_b$ and $\chi_c$
has the same origin as that in pure La$_{2}$CuO$_4$. This means that
$\chi_b(T)$ and $\chi_c(T)$ are still governed by the canted moments at
$x$=0.02 and 0.03 and reflect the growth of $\xi_{\rm AF}$. It is useful
to note that the DM interaction, in which the $b$ and $c$ axes are
equivalent, alone cannot account for the observed anisotropy between
$\chi_b$ and $\chi_c$ at high temperatures.

Another evidence for the spin alignment to remain anisotropic in samples
without the N\'{e}el order is that $\chi_b$ becomes {\it smaller} than
$\chi_a$ [Figs. 2(a) and 2(b)] at high temperatures. Again, the DM term
alone cannot account for the observed crossing of $\chi_a(T)$ and
$\chi_b(T)$ within the picture of 2D Heisenberg antiferromagnets,
because the contribution of the canted moments (which is confined in the
$bc$ plane) results in $\chi_b > \chi_a$. While a $T$-independent shift
of (1-1.5)$\times$10$^{-7}$ emu/g between $\chi_c$ and $\chi_a$ may be
associated with the anisotropic Van Vleck contribution \cite{SG2}, the
small ($\sim 3^{\circ}$) tilt of the oxygen octahedra in the $bc$ plane
of LSCO \cite{review} cannot generate a comparable ($\sim
0.4\times$10$^{-7}$ emu/g) difference between $\chi_a$ and $\chi_b$. On
the other hand, this difference can be readily understood if we assume
that the spin alignment at this doping is still governed by the easy
spin $b$-axis and hard $a$-axis; in this case, $\chi_a$ ($\chi_b$)
reflects the transverse (longitudinal) AF susceptibility similar to that
of a conventional antiferromagnet and thus $\chi_a > \chi_b$ is
naturally expected. It is worth noting that $\chi_a$ at moderate
temperatures is essentially independent of both temperature and doping
(the absolute magnitude of $\chi_a$ is always around 1$\times$10$^{-7}$
emu/g at moderate temperatures); this is naturally expected if $\chi_a$
reflects the transverse spin susceptibility of the AF state regardless
of doping.

Given the unusual anisotropy of $\chi$ at high temperatures, it is
intriguing to see what happens at low temperatures in the ``SG'' state,
which actually turns out not to be the canonical isotropic SG state
conjectured from studies of twinned crystals \cite{SG2}. Let us first
discuss the origin of the moments that are responsible for the
susceptibility upturn at low temperatures (and hence for the SG
behavior). One can immediately see from Fig. 2 that the low-temperature
susceptibility is notably anisotropic. The anisotropy between $a$ and
$bc$ apparently comes from the canted moments, which add extra moments
to $\chi_b$ and $\chi_c$, and therefore a simple Curie-Weiss analysis is
applicable only to $\chi_a(T)$; for $x=0.03$, we can estimate that only
$\sim$0.22\% of the total Cu spins participate in the Curie-Weiss
susceptibility. The origin of this Curie term in $\chi_a$ is expected to
give a clue to understanding the unusual magnetic state. One can see
that this Curie term persists even in the N\'{e}el state [see Fig.
1(c)], where there should be no free moments along the $a$-axis, and it
increases systematically with doping (which speaks against the
magnetic-impurity origin). The spin texture \cite{texture} is not likely
to be the origin of the Curie term, because spins in the AF domains
appear to be always aligned transverse to the $a$-axis. On the other
hand, if some spins are not engaged in the AF order, they may produce
the Curie term in $\chi_a$; for example, the spins located at the
frustrated borders between AF domains may actually behave as free spins.
In this case, $\chi_b$ and $\chi_c$ contain contributions from both the
free spins (on the boundary) and the canted moments (in the AF domains),
and thus should be larger than $\chi_a$.

\begin{figure}[!t]
\vspace{5pt}
\leftskip6pt
\epsfxsize=0.9\columnwidth
\centerline{\epsffile{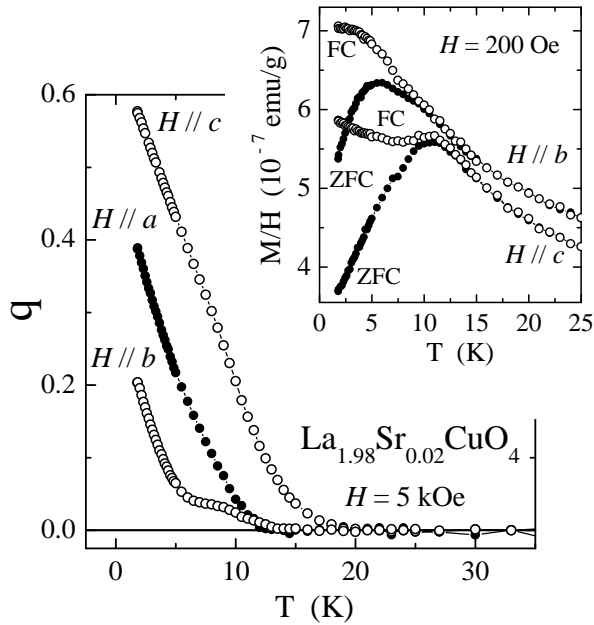}}
\vspace{5pt}
\caption{Anisotropic spin freezing in LSCO with $x$=0.02; $q$ is the
``spin-glass'' order parameter defined by
$\chi(T)=\chi_0+(1-q)C/(T+\Theta)$ \protect\cite{SG2}. Inset:
Low-temperature magnetization taken on heating after zero-field cooling
({\large$\bullet$}) and cooling at $H=200$ Oe ({\large$\circ$}). Note
the difference in the peak temperature between $b$ and $c$.}
\label{fig3}
\end{figure}

Whatever the origin of the magnetic moments contributing to the
low-temperature susceptibility is, they all freeze upon cooling below
some temperature $T_{SG}$, as is manifested in the steep decrease of
$\chi$ measured after zero-field cooling (inset of Fig. 3). However, the
freezing phenomenon turns out to be much more complicated than was
considered before \cite{SG2}. As can be seen in Fig. 3, the SG order
parameter $q$ (defined in the caption) starts to grow from different
temperatures for all three directions, indicating $T_{SG}$ is
anisotropic; also, the position of the peak in $\chi(T)$ clearly depends
on the field direction. Note that the anisotropy is not restricted to
the difference between $\parallel c$ and $\perp c$ directions, but is
well pronounced within the plane as well. We note that this anisotropic
freezing is essentially reproduced in the $x$=0.03 sample and thus is
robust against the change in doping.

We can list two possibilities that cause the unexpected susceptibility
anisotropy observed in the SG regime of the phase diagram. One is a
strong Ising-like spin anisotropy which aligns spins along the $b$-axis
at temperatures much above $T_N$. When this Ising-like anisotropy is
combined with a large $\xi_{\rm AF}$, there will be a domain structure
with the AF domains separated by the antiphase boundaries. Another
possibility is a new magnetic order, such as a unidirectional SDW state
\cite{1Da,1Db,stripe}, which may emerge upon suppression of the N\'{e}el
state. Both the domain structure and SDW bear a lot of similarities and
can readily account for the in-plane magnetic anisotropy. While more
studies are certainly necessary to arrive at a conclusive picture, the
magnetic state in lightly-doped cuprates appear to be significantly
different from a simple 2D Heisenberg antiferromagnet.


\begin{references}
\bibitem{note}
The homogeneous (phase-separated) state
has its own $T_N$ of 240 K (300 K),
which can be estimated by fitting $\chi_c(T)$. Thus,
$T_N$ becomes effectively $T$ dependent and increases upon
cooling, which can explain the observed sharpening of the
peak in $\chi(T)$.
\end{references}
\end{document}